# Convolutional Networks for Image Processing by Coupled Oscillator Arrays

Dmitri E. Nikonov, Ian A. Young, and George I. Bourianoff

Components Research, Intel Corp., Hillsboro, Oregon 97124 USA

Abstract

A coupled oscillator array is shown to approximate convolutions with Gabor filters for image processing tasks. Pixelated image fragments and filter functions are converted to voltages, differenced, and input into a corresponding array of weakly coupled Voltage Controlled Oscillators (VCOs). This is referred to as Frequency Shift Keying (FSK). Upon synchronization of the array, the common node amplitude provides a metric for the degree of match between the image fragment and the filter function. The optimal oscillator parameters for synchronization are determined and favor a moderate value of the Q-factor.

## 1. Introduction

Digital computing has achieved an unprecedented success in the last four decades due to scaling of complementary metal-oxide-semiconductor (CMOS) transistors [1] according to the Moore's law [2]. It enabled personal computing, the Internet, mobile communications, wearable devices, and the "Internet of Things", in addition to a profound change in almost any area of human culture. Meanwhile analog computing progress was not nearly as impactful and was mostly confined to technology niches, see e.g. [3]. One particular type of analog computing – neuromorphic computing was continually pursued, see e.g. reviews [4,5]. It is based on a profound observation: though computers are much better than humans in some tasks (like arithmetic operations with large numbers), humans handily beat computers in other tasks (like recognition of complex objects with ambiguous or missing details). Even though software implementations of such tasks exist, they are unreliable, and take too much time and energy compared to customer needs. Thus the goal of neuromorphic computing is to emulate, to various degrees, information processing occurring in the human brain. The neuromorphic thrust gave rise to various approaches, such as artificial neural networks [6], cellular neural networks [7], LEGION [8], oscillator computing [9], optical computing [10], holographic computing [11], etc. They have been applied with various degrees of success to problems such as pattern or image recognition, content addressable memory, and Bayesian inference.

In this work we explore one type of such computing – coupled oscillator array (COA) as envisioned in works by Hoppensteadt and Izhikevich [12], Itoh and Chua [13], Corinto, Bonnini, and Gilli [14]. Most of the previously published schemes are based on Phase-Shift Keying (PSK). In PSK, the patterns are encoded as the phase of a coupling coefficient between oscillators and the outputs are obtained from relative phases of oscillators. In this article we are focusing on an alternative approach - Frequency Shift Keying (FSK) [15]. In FSK, the patterns are encoded by the frequency shifts of the oscillators, and the matching decision is obtained from the amplitude of the sum of outputs from all oscillators, which in its turns depends on synchronization of oscillators. The FSK method will be more fully explained in this paper.

Previously Couple Oscillator Array (COA) schemes were applied to matching binary vectors, which commonly result from extraction of features from images. Matching an input feature vector to one of the memorized feature vectors thus allowed one to recognize and classify images (or their fragments). However most of the computational load in image recognition comes from the image pre- processing for object detection.

Presently the tasks of image preprocessing and object detection are accomplished by digital algorithms such as convolutional neural networks (CNN) [16,17],or HMAX [18,19]. In this work we demonstrate that a COA can perform an operation which approximates a convolution. More importantly, we argue that it is more energy efficient than the digital counterpart. The paper is organized as follows. In Section 2 we review the Gabor filters and the task of their convolutions with an image. The numerical model of the oscillator array is described in Section 3. The examples of oscillator dynamics are presented in Section 4. The dependence of oscillator synchronization on parameters is explored in Section 5 and a surprising optimum value of the Q-factor is found. The conclusions are formulated in Section 6.

## 2. Image Processing with Gabor Filters

An important step in image processing and recognition is the convolution of the image with selected filter functions for feature extraction. The algorithms of convolutional neural networks include consecutive stages of filtering and pooling resulting patterns. An important class of filters is Gabor [20] filters. They correspond to determination of oscillatory changes in images in a certain direction, and edge detection as a special case of this. They are used to detect the edges, orientation and periodicity of selected objects. Typically they are defined by functions containing a product of a localized part (such as a Gaussian) and an oscillatory part (such as a plain wave given by a complex exponent):

$$G(x, y) = \frac{1}{2\pi S^2} \exp\left(-\frac{x^2 + y^2}{2S^2}\right) \exp(i 2\pi k (x\cos\theta + y\sin\theta)), \qquad (1)$$

The imaginary part of the resulting function is typically used to determine image directionality and periodicity. To apply such filters to pixelated images, this function is taken at discrete, integer values of spatial coordinates $x, y$. In this work we choose 7x7 blocks of pixels, and the set of 24 Gabor filters [21] given by the following parameters: $S = 3$, the wavevector amplitudes $k = 0.2; 0.4; 0.6; 0.8$, and the wavevector angles $\theta = 0°, 30°, 60°, 90°, 120°, 150°$. The coordinates $x, y$ have the origin at the center pixel. The resulting set of filter function values is shown in Figure 1. The colors of each square expresses the interpolation of the values of the four pixels in its corners.

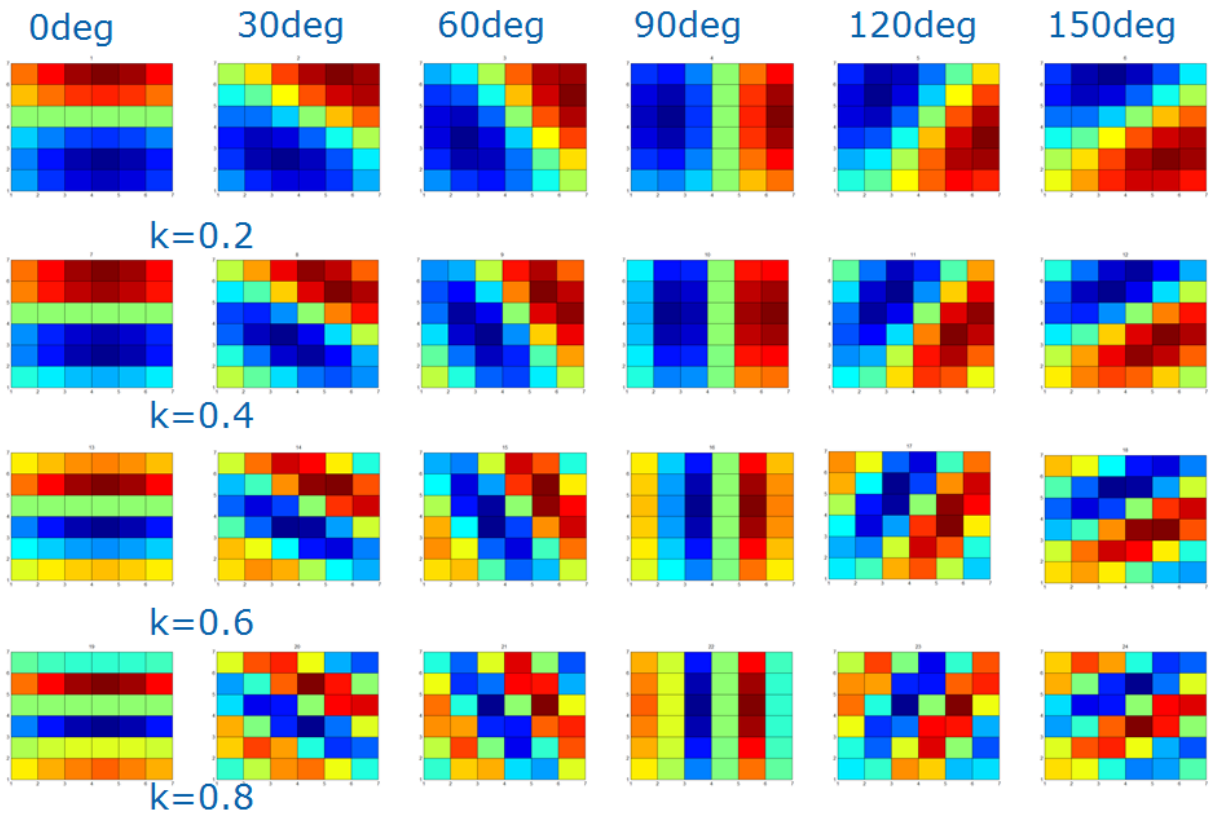

**Figure 1. A set of 24 Gabor filters of the size 7x7 used in this work. The colors of the rectangles designate the average value of the 4 surrounding pixels and correspond to the range from -1 (dark blue) to +1 (dark red).**

The filtering operation is performed as a convolution of each filter $G$ with consecutively all fragments $F$ of the image of the same size.

$$(F \cdot G)(x', y') = \int F(x'-x, y'-y)G(x, y)dxdy. \qquad (2)$$

For pixelated images, the convolution becomes a matrix product

$$(F \cdot G)_{kl} = \sum_{i,j=1}^{n'} F_{k-i,l-j}G_{ij}, \qquad (3)$$

where the summation is performed over the block of pixels n' in one dimension. This is equivalent to a dot product (albeit with a decreasing index for one of the factors)

$$F \cdot G = \sum_{i=1}^{n} F_{n-i}G_i. \qquad (4)$$

We illustrate the filtering operation by the example of a convolution of a grayscale image "Lenna" [22] with one of the above Gabor filters in Figure 2. As a result, the edges at a certain angle ($-60^o$ in this case) are highlighted. Bright to dark change of intensity is mapped to a bright edge (positive values), and the dark to bright change of intensity is mapped to a dark edge (negative values). The majority of the area of the image is gray (corresponding to values close to 0).

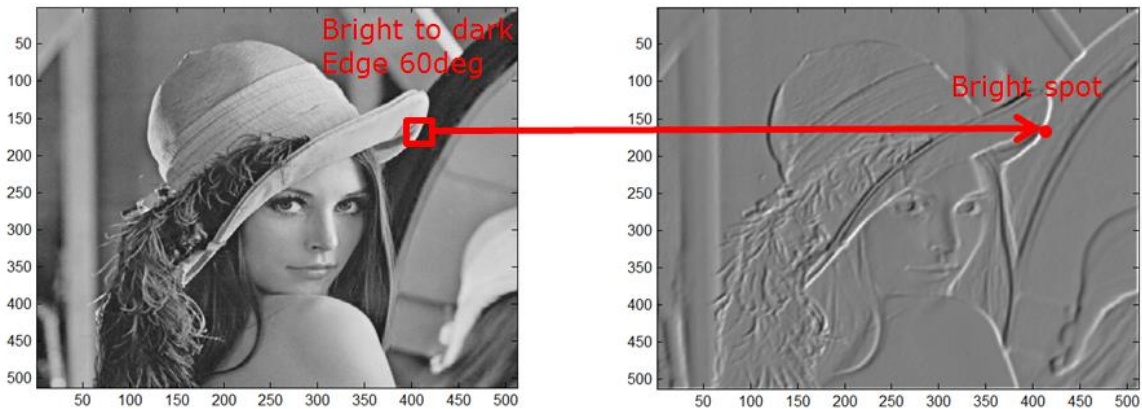

**Figure 2. A grayscale image "Lenna" and its convolution with Gabor filter #9 shown in Figure 1.**

In general, a dot product between vectors

$$d_{kl} = (x_k, x_l) = \frac{1}{n}\sum_{i=1}^{n} \xi_{k,i}\xi_{l,i}, \qquad (5)$$

is related to the Euclidean distance between them

$$\|x_k - x_l\|^2 = (x_k - x_l, x_k - x_l), \qquad (6)$$

by the following relationship (for vectors normalized to unity)

$$\|x_k - x_l\|^2 = 2(1 - d_{kl}). \qquad (7)$$

Thus the problem of the convolution with filters is equivalent to the problem of finding the Euclidean distances between the vectors corresponding to the fragment and a filter.

We picked an arbitrary fragment from the image (265:271,270:276) and calculated its dot products with the above 24 Gabor filters. The results are shown in Figure 3. It shows that four filters (#3,4,9,10 numbered by rows in Figure 1) have the largest values of the dot product and therefore are closest filter functions to the fragment.

| 1 | 1.95 | 13 | 0.88 |
|---|---|---|---|
| 2 | 5.11 | 14 | 2.08 |
| 3 | 7.14 | 15 | 4.30 |
| 4 | 7.25 | 16 | 4.12 |
| 5 | 5.20 | 17 | 3.10 |
| 6 | 1.73 | 18 | -0.31 |
| 7 | 2.18 | 19 | -0.28 |
| 8 | 5.22 | 20 | 0.14 |
| 9 | 8.04 | 21 | 1.18 |
| 10 | 8.44 | 22 | -0.02 |
| 11 | 5.85 | 23 | 0.91 |
| 12 | 1.29 | 24 | -0.93 |

**Figure 3. Values of the un-normalized dot product of an arbitrarily taken fragment of the image and the 24 Gabor filters. Four closest matches are highlighted.**

Both the fragment and the Gabor filters are vectors in a multi-dimensional (49-dimensional in this case) space. To visualize the relative positions of the vectors in such a multi-dimensional space, it is often helpful to apply the principal component analysis (PCA) [23]. It is equivalent to a projection of the multi-dimensional space on a two-dimensional subspace defined by the principal components PC1 and PC2, which are selected such that the spread of the set of vectors is maximal. PCA results are shown in Figure 4. They are consistent and supplementary to the one given by dot products: the above 4 filters are seen to be close to the fragment.

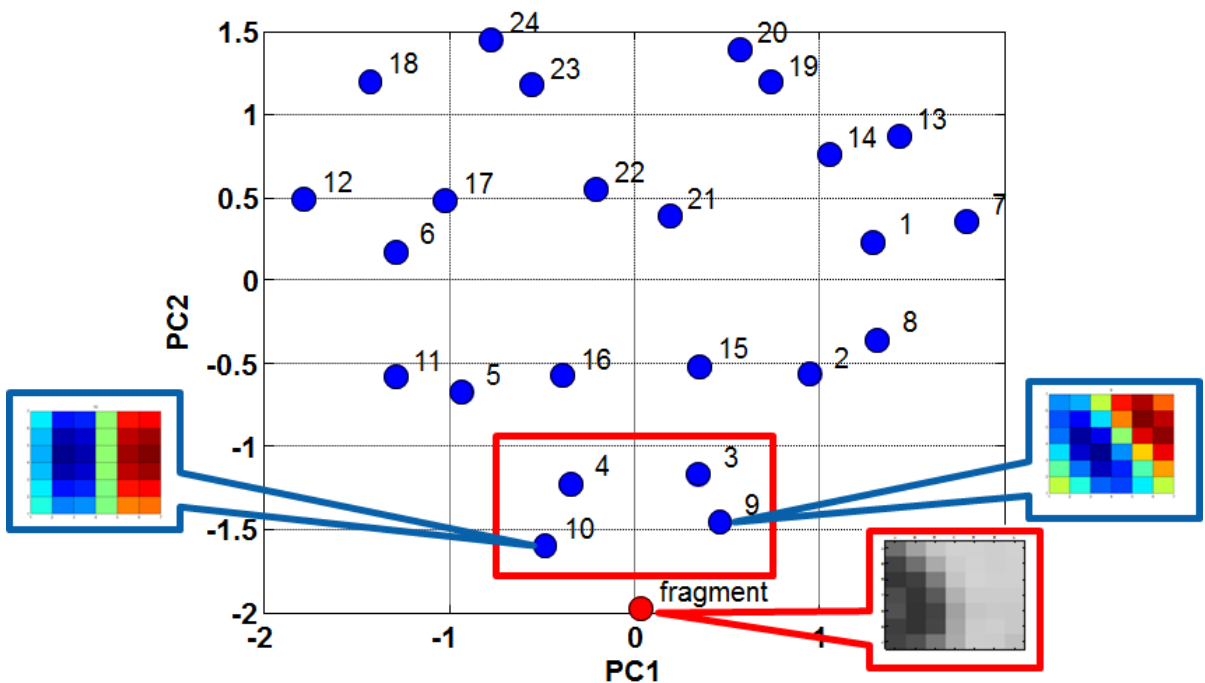

**Figure 4. Projections of the image fragment and the 24 Gabor filters on the plane of two principal components.**

Determining the Euclidean distance between real-valued vectors by performing floating-point calculations is straightforward but computationally intensive. Moreover, the 32 or 64 bit precision inherent in general purpose computing is often not required for image recognition. In this work we explore an approach of approximate analog, non-Boolean computation of a quantity approximating the dot product. This approach is based on synchronization of a coupled oscillator array and measuring the output voltage at a summing node. To explain the motivation for this approach, we compare the energy required for such calculation. A present day microprocessor typically dissipates 3nJ per FLOP. With the length of the vector of n=49 the required energy is $E_{MPU} \sim 150nJ$ per FLOP. Let us assume that a COA using spin torque oscillators [15] can perform this computation in 1.2ns with a current of $100\mu A$ and voltage of 0.1V. Then the expended energy will be $E_{COA} \sim 0.6pJ$. Therefore the non-Boolean computation can be orders of magnitude more efficient than the conventional digital calculation. The actual number of FLOPS required to calculate an Euclidian distance metric is much larger than 1 so this comparison is a very conservative estimate.

### 3. Model of a Coupled Oscillator Array

We apply an array of coupled oscillators to the task of recognition and seek a design where synchronization of oscillators would correspond to a match. Various -topologies of the oscillator array can be -considered. One can use CMOS based oscillators, such as ring oscillators [24] or phase-locked loops [25,26]. Alternatively, one can use spin-torque oscillators (STO) [27] sharing a common ferromagnetic free layer. STOs are sub 100 nanometer-scale devices, acting as compact microwave oscillators [28,29,30,31,32,33]. The self-sustaining oscillations are generated by the flow of spin-polarized currents (spin-torque) into a thin magnetic layer and the magnetization oscillations can be detected by the resistance change in the same current path [34]. The magnetic oscillations may create propagating spin waves [35,36] in the ferromagnetic free layer, which provide a non-electrical coupling mechanism between STOs [37]. In general, rigorous modeling of the behavior of these oscillators is computationally intensive [38]. In this work we will aim to gain insights into the COA operation by using simple models of oscillators. Certainly the question remains of how faithfully these simple models reflect behavior of realistic oscillators. We speculate that simple models contain the essential features – non-linearity and the limit cycle – pertinent for realistic oscillators.

One such model is the modified van der Pol equation. This model closely follows those of previous publications, e.g. [12].

$$\frac{dz_i}{dt} = (\rho + i\omega_i)z_i - \rho z_i |z_i|^2 + \varepsilon \sum_{j=1}^{n} z_j, \tag{8}$$

where $z = x + iy$ is the complex amplitude of an oscillator, $\rho_i$ is the damping parameter, $\omega_i$ is the cyclic frequency of oscillation, $\varepsilon$ is the strength of coupling between oscillators, the sum is performed over $n$ of oscillators comprising the array. These equations exhibit oscillations containing a limit cycle with the amplitude of 1. The number of oscillators is set by the dimension of the patterns to be compared. We use random uniformly distributed phases and Gaussian distributed amplitudes with the variance of 1 as the initial conditions. These equations describe equal coupling of each oscillator to each. In realistic circuits, the coupling coefficients will be different due to the signal delay and attenuation as well as noise. These factors are planned to be addressed in future publications. Such coupling does not necessitate $n(n-1)/2$ physical connectors. In fact, all oscillators can be connected to a common signal node, i.e. an "Averager" by just $n$ physical connectors (star coupling), see Figure 5. Its purpose is to sum outputs of all oscillators and then to broadcasted it back to inputs of the oscillators. In electronic circuits, the signals could be AC voltages. The "Averager" has a convenient property that its signal amplitude stands for the quantity which, as will be shown, represents the degree of match (DOM)

$$DOM(t) = \frac{1}{n} \left| \sum_{j=1}^{n} z_j(t) \right|, \tag{9}$$

Previous schemes [12,13,14] use variable phases of coupling between the oscillators to encode patterns; i.e. phase-shift keying (PSK). In contrast, we use fixed phase of coupling coefficients and encode the patterns via frequencies of the oscillators, i.e. frequency-shift keying (FSK), in the following manner. The oscillators must be voltage controlled ones (Voltage Controlled Oscillators – VCOs) and have their initial frequencies shifted from the center frequency $\omega_0$ by the DC control voltage inputs proportional to the difference of the fragment $F$ and the Gabor filter $G$ patterns on a pixel by pixel basis :

$$\omega_j = \omega_0 + \Delta\omega(F_j - G_j), \tag{10}$$

The characteristic shift value is $\Delta\omega$, and the index $j$ labels the pixels in the patterns and VCOs.

As the oscillators evolve, their instantaneous frequencies drift. If they synchronize, their final frequencies lock. They can also partially synchronize, with some oscillators remaining at different frequencies. The quality factor of the VCOs is defined as $Q = \rho/\omega_0$.

To put this model in perspective, let us compare it with the Kuramoto model [39], which has been widely used to represent arrays of coupled oscillators. The Kuramoto model contains only phases of oscillators. In contrast, all our models involve both amplitudes and phases of oscillators and thus are more general. In addition, manipulating and measuring phase shifts is inherently more complicated than manipulating and measuring frequency shifts.

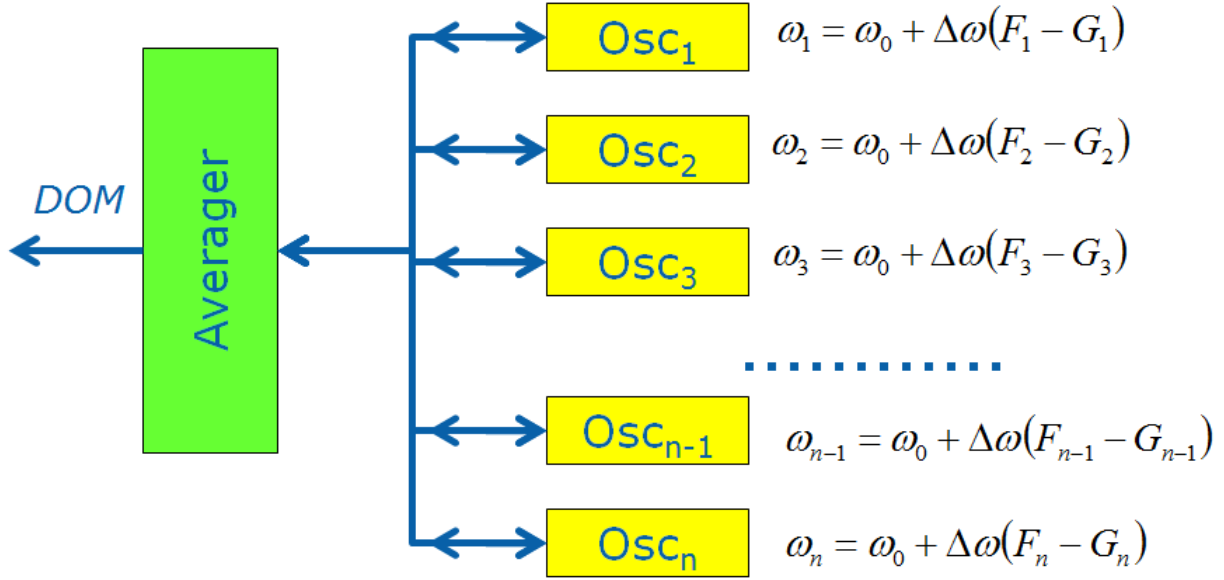

**Figure 5. Block diagram of COA with FSK. Signals are passed in both directions through links.**

## 4. Examples of Dynamics of the Oscillator Array

Next we solve equations (8) to simulate a few examples of COA evolution to demonstrate the nature of synchronization. If not otherwise specified, the coupling strength $\varepsilon = 0.002$, the damping $\rho = 0.03$, and the frequency spread $\Delta\omega = 0.2$ in the units of the center frequency $\omega_0$.

Figure 6 depicts the case of bad match between the fragment and the Gabor filter #12. Consequently, the oscillators do not synchronize with each other. In this case most of the oscillators have their initial frequencies shifted away from the center frequency by a value exceeding the locking range set by the coupling strength. Therefore the relative phases shown in Figure 6(a) continue to grow with time in an approximately linear manner (because the frequencies are different as shown in Figure 6(b)). Instantaneous frequencies shown in Figure 6(b) oscillate approximately within their original spread range and do not converge to a common value. Consequently, the oscillator outputs interfere destructively the DOM variable oscillates around a value significantly smaller than 1.

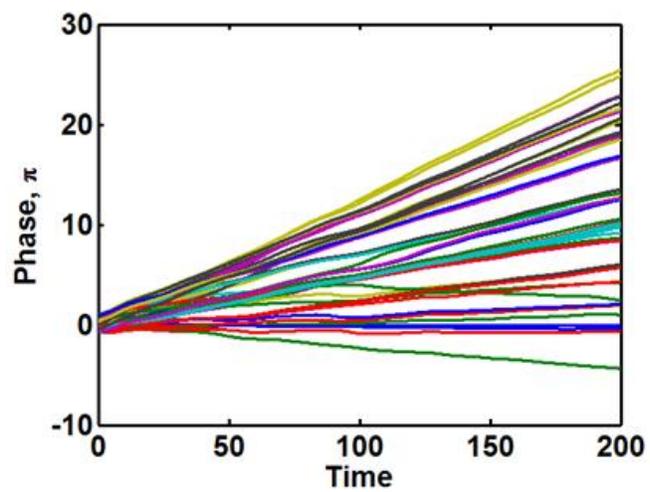

a)

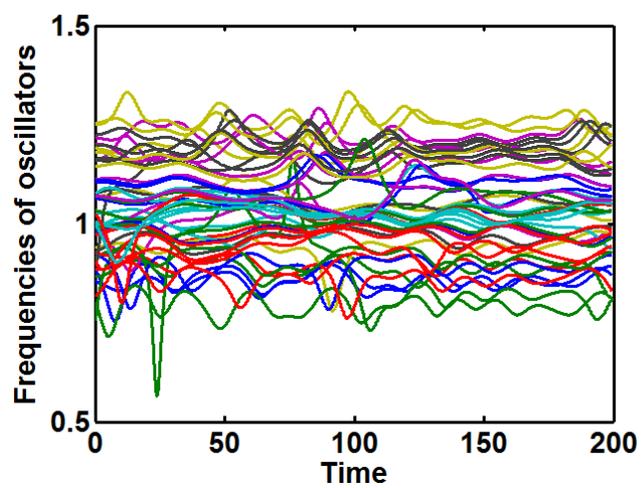

b)

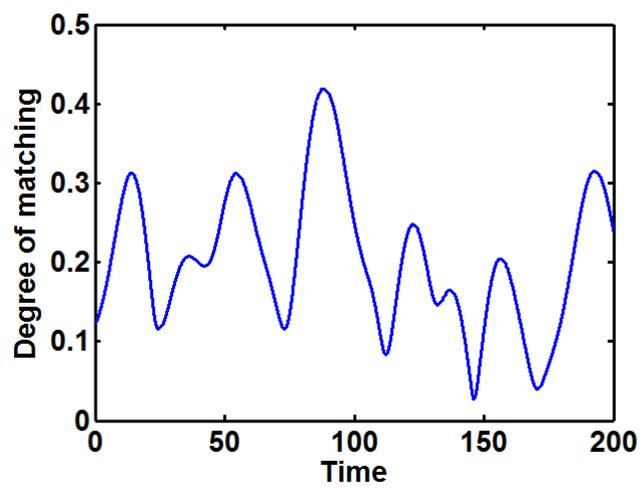

c)

**Figure 6. Evolution of COA in case of a bad match of the fragment to the Gabor filter (#12). a) .Oscillator phases vs. time. b) Oscillator instantaneous frequencies vs. time. c) Averager amplitude (i.e. DOM) vs. time.**

In contrast, Figure 7 depicts the case of good match between the fragment and the Gabor filter #9 where the oscillators do synchronize with each other. In this case most of the oscillators have their initial frequencies shifted away from the center frequency within the locking range. Therefore the relative phases (a) tend to a constant value at long times. All instantaneous frequencies (b), after some oscillations, lock to a value close to the center frequency. Consequently, the oscillator outputs mostly add in phase and the DOM variable grows to a weakly oscillating value close to 1.

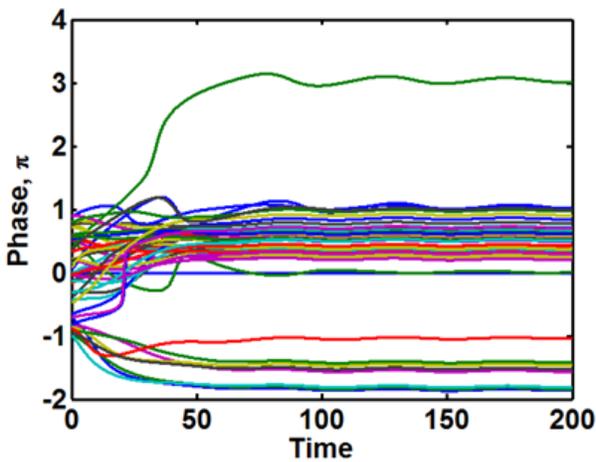

a)

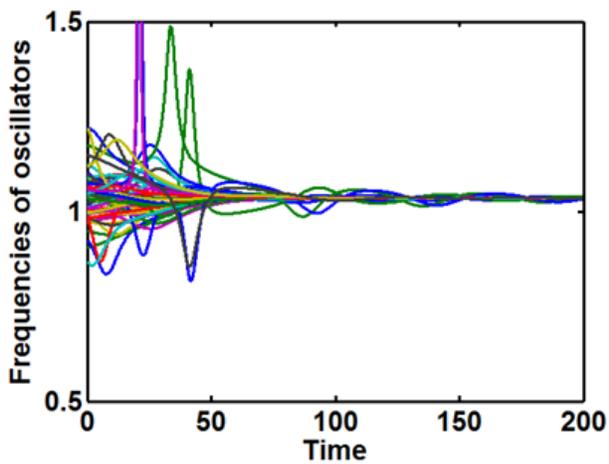

b)

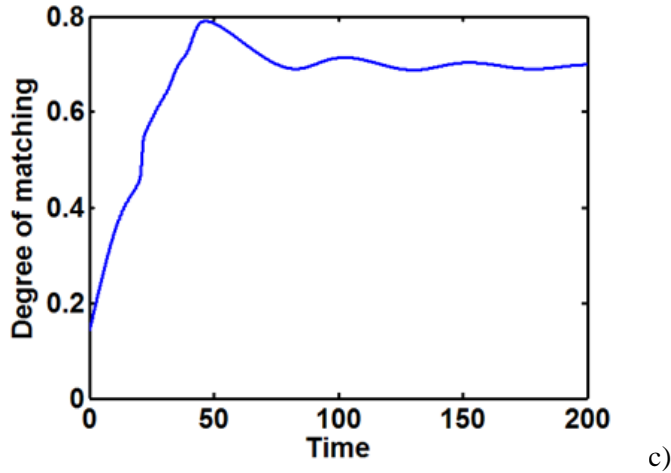

c)

**Figure 7. Evolution of COA in case of a good match of the fragment to the Gabor filter (#9). a) .Oscillator phases vs. time. b) Oscillator instantaneous frequencies vs. time. c) Averager amplitude (i.e. DOM) vs. time.**

To average over the COA oscillations one can integrate the DOM signal over a certain period of time. We choose the duration of 15 oscillation periods here. If one is required to compute DOM for the set of filters simultaneously, one can build a set of COA, one each for comparing each fragment fragment with one of the filters. After that, the resulting DOM signals can be processed by a k-winner-take-all module which determines the (several) 'winners', i.e. matches with the highest DOM.

By performing simulations with various Gabor filters we demonstrate that DOM is indeed a good proxy for the dot product. Figure 8a depicts the simulated DOM vs. the number of the Gabor filter. As expected, the higher DOM corresponds to filters #3,4,9,10. The correspondence between the exact dot product and DOM is shown in Figure 8a. It demonstrates a clear correlation between DOM and the dot product. The dependence is non-monotonic and noisy. It varies slightly between runs and for various fragments of an image. This is caused by the randomness of the initial conditions. Note also a certain non-linearity in the dependence. In other words –a larger dot product is accompanied by a higher slope of DOM relative to the dot product. This attractive feature amplifies the characteristic of a few winners of the match. One can speculate that it is caused by the non-linear character of synchronization.

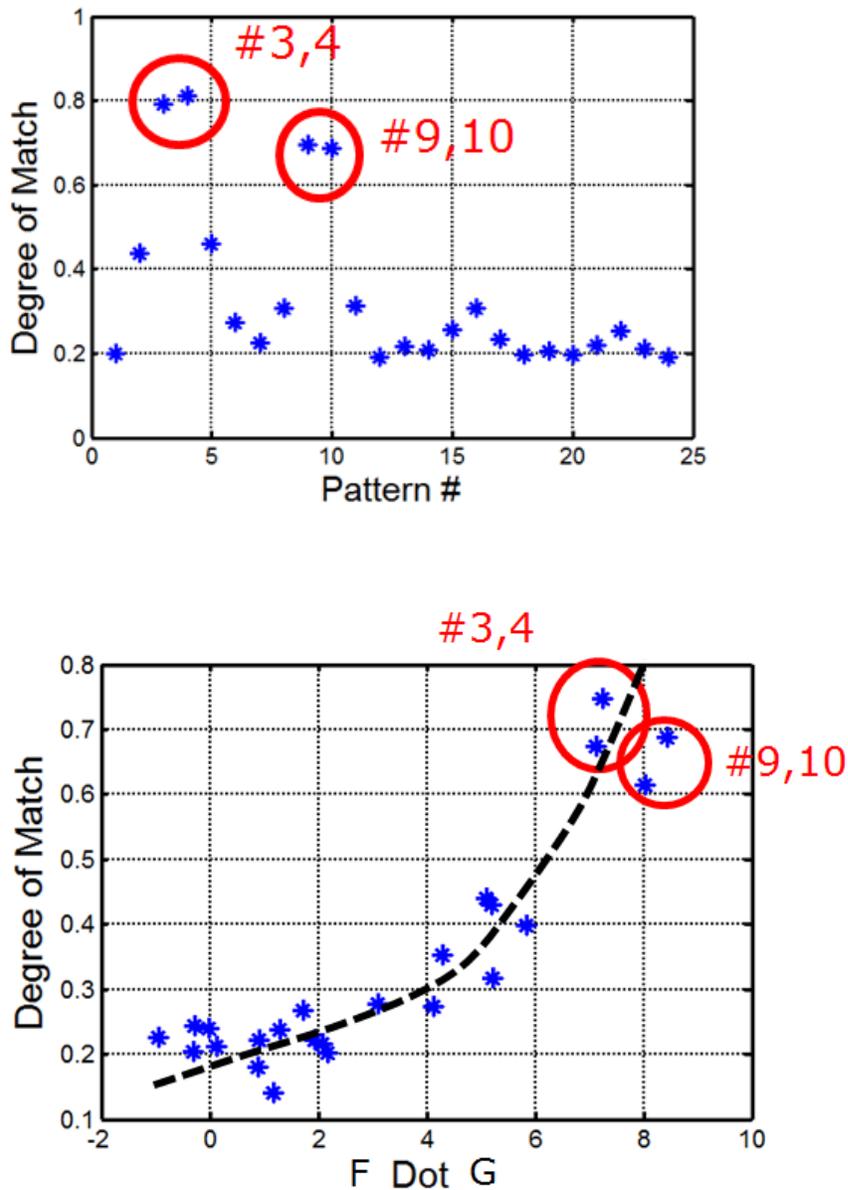

Figure 8. a) Degree of match for the 24 Gabor filters. b) Degree of match vs. the dot product of the fragment and the filters.

Another insight to the nature of synchronization can be gained by examining the spectrum of frequencies. As was shown above, a good match and synchronization corresponds to locking of instantaneous frequencies to one value, while a bad match corresponds to a wide spread of frequencies. Figure 9 depicts the histogram of frequencies for various Gabor filters. As we can see, in the four cases of good match, all the frequencies end up in one bin of the histogram. For other cases of worse match, the frequencies are distributed over several bins. The frequency spectrum serves as a convenient indicator for optimizing parameters of the array for better discrimination. In other words, one aims for a narrow frequency spectrum for only a few winners of the match and wider spectrum for the rest of the compared patterns.

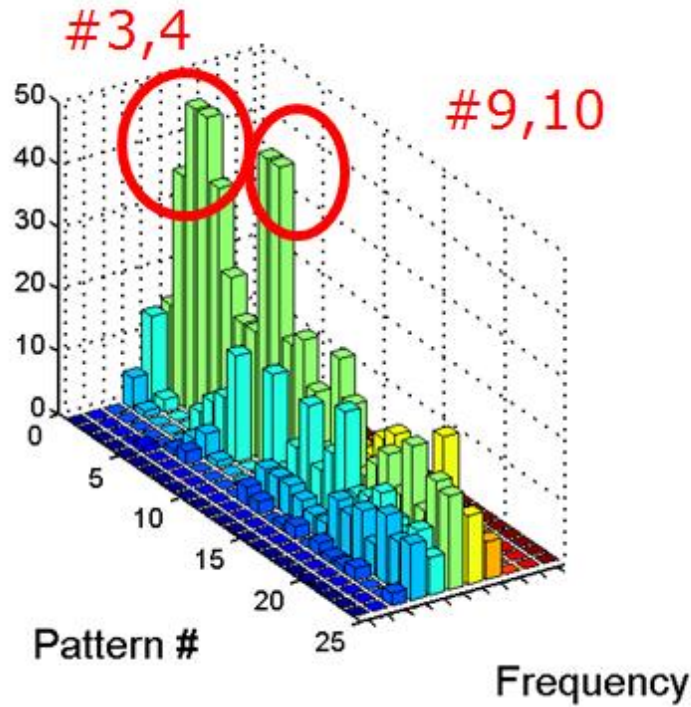

**Figure 9. Histogram of the frequency spectrum for the 24 Gabor filters.**

Yet another demonstration that the COA dynamics adequately emulates the dot product (and similarly convolution) is obtained from the comparison in Figure 10 of the direct convolution and DOM (actually computed for every second pixal column and row). As one can see, the images are in qualitative agreement: edges are found in the same locations for both. The DOM image is noisy due to the randomness of initial conditions mentioned above.

## 5. Synchronization Dependence on Parameters

Now we proceed to explore how the parameters of the VCO oscillator array affect the process of synchronization and its correspondence to the emulation of the dot product (and similarly convolution).

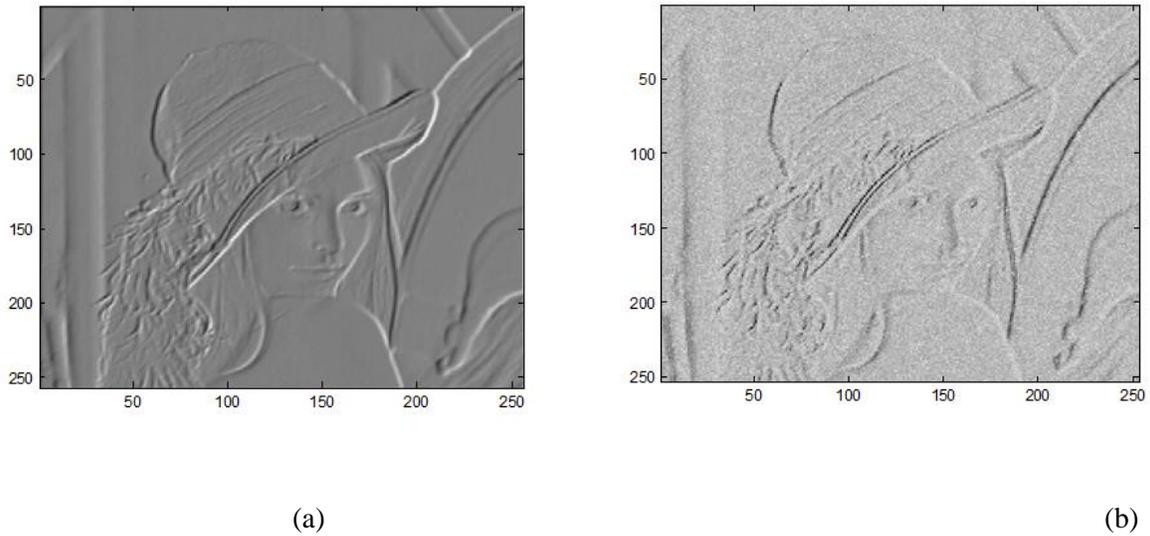

(a) (b)

**Figure 10. Direct convolution of the "Lenna" image (a) and the degree of match representing the convolution (b).**

The cases where coupling strength is varied are shown in Figure 11. At a lower value of oscillator coupling strength, the DOM is low for all Gabor filters, and the spectrum of frequencies is wide for all Gabor filters as well. In other words this coupling strength is insufficient to achieve good synchronization. At a higher coupling strength, DOM is high for all Gabor filters, and the frequencies are locked for all Gabor filters. In other words, synchronization is too strong and we lose the differentiation between the matching patterns. Only a range of intermediate values of coupling strength produces good differentiation between Gabor filters examined in the previous section. This range is still approximately 2x, and easy to satisfy even considering variations of the parameters cause by the fabrication process.

$\Delta\omega = 0.2$

$\varepsilon = 0.001$

Do not lock

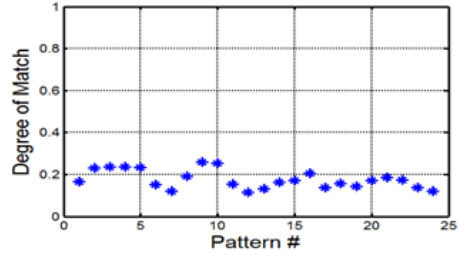
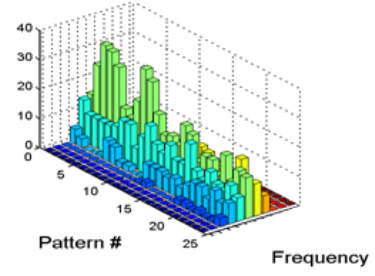

$\varepsilon = 0.002$

Good locking

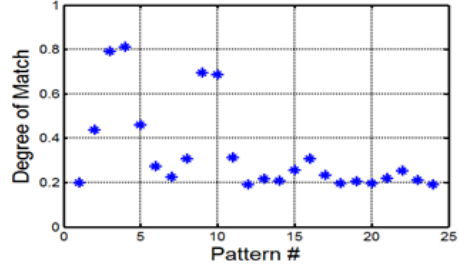
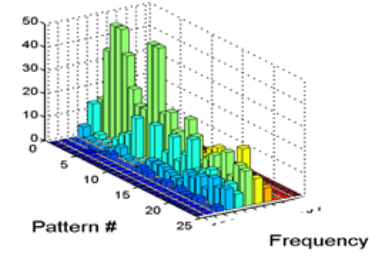

$\varepsilon = 0.004$

Too many lock

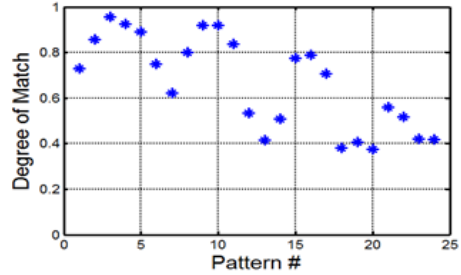
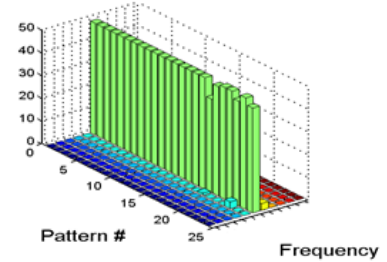

**Figure 11. Cases of varying coupling strength in a COA: the dependences of the degree of match on the Gabor filter number and the histograms of the final frequencies of the oscillators. The parameters are shown in units of the central cyclic frequency $\omega_0$.**

From the analysis of the above simulations, we obtain the following empirical relations. The optimal frequency spread is related to the individual coupling strength multiplied by the number of the oscillators:

$$\Delta\omega/\omega_0 \approx 2\varepsilon n. \qquad (11)$$

The number of oscillation periods required to achieve frequency locking is inversely proportional to the frequency spread:

$$T_{lock} \approx 2.4\omega_0/\Delta\omega \approx 1.2/(\varepsilon n). \qquad (12)$$

Finally we examine the influence of the oscillator Q-factor. It COA exhibits a trend of undesired high or low values of Q with the optimum at the intermediate values. This is similar to the dependence on the coupling strength. For small Q, some of the Gabor filters with good expected match have some of the oscillators not locked with the majority. For large Q, some of the Gabor filters with not the highest expected match still have perfect locking of all the COA oscillators. Therefore the differentiation of DOM deteriorates in both extremes of oscillator Q. The character of synchronization is not very sensitive to the value of Q. Even as Q is varied by 20x, the frequency spectrum remains qualitatively similar. A surprising observation is that the optimal oscillator Q is in the range of 20. This is much lower than targets for Q sometimes set in experimental efforts. In fact such high Q values may be detrimental. In contrast, relatively low Q values observed in CMOS ring oscillators or STOs may be sufficient.

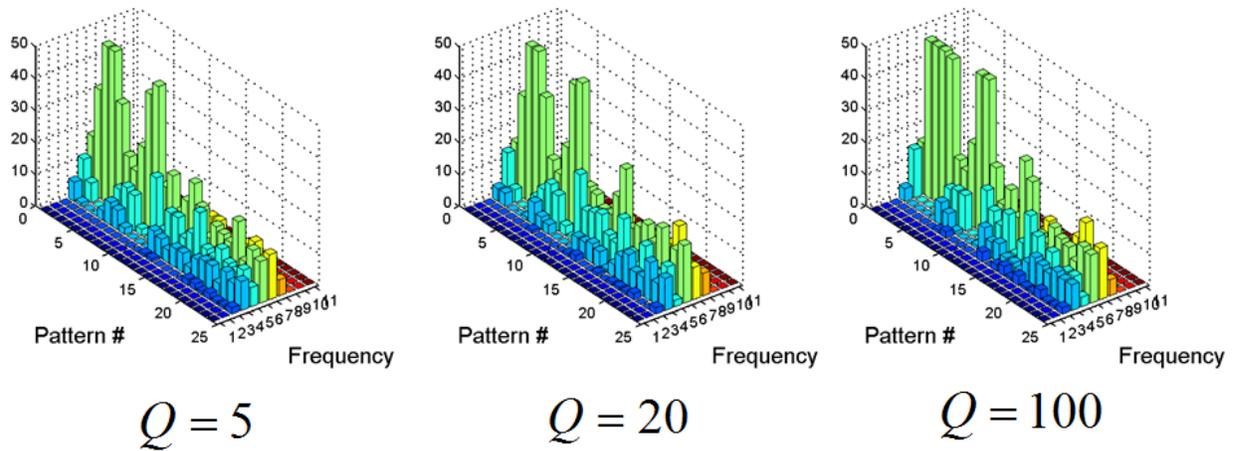

Figure 12. Cases of varying oscillator Q-factor: the histograms of the final frequencies of the oscillators.

## 6. Conclusions

In summary, we simulated the operation of a coupled oscillator array with frequency shift keying. In the array, each oscillator's frequency shift encodes the difference between an image fragment's pixel and a Gabor filter. We demonstrated that the synchronization of the oscillators in the COA is determined by the frequency shift, and if they synchronize, the frequencies converge to the same value. The amplitude of the signal on the common coupling node, the "Averager", determines the degree of match. DOM approximates the convolution of the fragment and the Gabor filter. Therefore we can replace an energy demanding floating point operation with an analog operation that has potentially 4 orders of magnitude smaller energy. Surprisingly, high values of Q-factor are not required for optimal synchronization and discrimination of the DOM. In the considered example a Q of approximately 20 is optimal.

## 7. Acknowledgements

The authors are grateful to Mike Mayberry and Dan Hammerstrom for helpful discussions.